\documentstyle[amssymb,prb,aps]{revtex}
\begin{document}
\twocolumn[\hsize\textwidth\columnwidth\hsize\csname
@twocolumnfalse\endcsname

\title{Hole-doping dependence of percolative phase separation\\
in Pr$_{0.5-\delta}$Ca$_{0.2+\delta}$Sr$_{0.3}$MnO$_3$ around half
doping}

\author{D. Niebieskikwiat,$^*$ R.D. S\'{a}nchez, L. Morales, and B. Maiorov}
\address{Comisi\'{o}n Nacional de Energ\'{\i}a At\'{o}mica-Centro At\'{o}mico Bariloche
and Instituto Balseiro, 8400 Bariloche, Argentina}

\date{\today}

\maketitle

\begin{abstract}

We address the problem of the percolative phase separation in
polycrystalline samples of
Pr$_{0.5-\delta}$Ca$_{0.2+\delta}$Sr$_{0.3}$MnO$_3$ for $-0.04\leq
\delta \leq 0.04$ (hole doping $n$ between $0.46$ and $0.54$). We
perform measurements of X-ray diffraction, dc magnetization, ESR,
and electrical resistivity. These samples show at $T_C$ a
paramagnetic (PM) to ferromagnetic (FM) transition, however, we
found that for $n>0.50$ there is a coexistence of both of these
phases below $T_C$. On lowering $T$ below the charge-ordering (CO)
temperature $T_{CO}$ all the samples exhibit a coexistence between
the FM metallic and CO (antiferromagnetic) phases. In the whole
$T$ range the FM phase fraction ($X$) decreases with increasing
$n$. Furthermore, we show that only for $n\leq 0.50$ the metallic
fraction is above the critical percolation threshold $X_C\simeq
15.5\%$. As a consequence, these samples show very different
magnetoresistance properties. In addition, for $n\leq 0.50$ we
observe a percolative metal-insulator transition at $T_{MI}$, and
for $T_{MI}<T<T_{CO}$ the insulating-like behavior generated by
the enlargement of $X$ with increasing $T$ is well described by
the percolation law $\rho ^{-1}=\sigma \sim (X-X_C)^t$, where $t$
is a critical exponent. On the basis of the values obtained for
this exponent we discuss different possible percolation
mechanisms, and suggest that a more deep understanding of
geometric and dimensionality effects is needed in phase separated
manganites. We present a complete $T$ vs $n$ phase diagram showing
the magnetic and electric properties of the studied compound
around half doping.

\end{abstract}

\vskip2pc] \narrowtext

\section{INTRODUCTION}

What is the origin of the ``Colossal Magnetoresistance" (CMR)?
Since the discovery of the CMR, an extensive research effort was
done in this subject in order to elucidate the reasons for the
appearance of an enormous reduction of the resistivity ($\rho$)
with the application of a magnetic field ($H$) in the manganese
perovskites A$_{1-x}$A$'_x$MnO$_3$ (A=La, Pr, Nd, Y and A$'$=Ca,
Sr, Ba) and in other related half
metals.\cite{jin94,hwang96,coey98,srfemo} However, the complexity
of the phase diagrams of these compounds indicate that the
underlying mechanism of the CMR is not universal. In addition, the
single- or poly-crystalline character of the samples also
influences the magnetotransport response of the material. In
strongly ferromagnetic (FM) compounds like
La$_{2/3}$Sr$_{1/3}$MnO$_3$, the single crystals exhibit the CMR
at the vicinity of the Curie temperature $T_C$ as a linear
decrease of $\rho$ with $H$ associated to the reduction of spin
fluctuations, while polycrystals show a very different behavior,
with a large fall of $\rho$ at low fields (LFMR) followed by a
much slower decrease of the resistance.\cite{hwang96} The LFMR is
related to the reduction of the missalignment of the magnetization
of contiguous grains and is enhanced with decreasing $T$. On the
other hand, a much higher decrease of several orders of magnitude
of $\rho$ is achieved in charge-ordered antiferromagnetic (CO/AFM)
compounds like
(Pr,Nd)$_{1/2}$Ca$_{1/2}$MnO$_3$.\cite{tokura99,tokunaga98} In
this case the CMR is related to the transformation of the CO/AFM
state into the FM metallic one, however magnetic fields of several
tens of Teslas are necessary to observe this effect.

In the recent past, an interesting and potentially useful
phenomenon of phase separation (PS) have been
discovered.\cite{dagottoreview,uehara99,moreo00,dario01a,dario01b,ritter00,huang00,babushkina00,hardy01,kim00,podzorov00,mayr01,xiong01}
The PS consists of the simultaneous coexistence of two or more
magnetic (electronic) phases, like the FM, CO/AFM and paramagnetic
(PM) ones. In phase separated manganites, the metal-insulator (MI)
transitions occur in a percolative way, and a high CMR is achieved
with moderated fields. With the application of a magnetic field an
increase of the FM clusters is favored and the resistivity
decreases due to the improvement of the geometrical factor for the
current transport. Several workers have started to address the
problem of percolation of the binary metal-insulator mixture, such
as critical percolation thresholds and critical exponents, and its
relation to
CMR.\cite{babushkina00,hardy01,kim00,podzorov00,mayr01,xiong01}
However, the wide range of possibilities given by the phase
diagrams of low-bandwidth manganites makes these works just the
beginning of the subject. For example, the major part of that
previous research was done near doping $3/8$, where CO appears at
high $T$ (but lower than $T_{CO}$) and the ground state is
essentially FM metallic. On the other hand, in compounds
presenting PS near half doping, as for example
Pr$_{0.5}$Ca$_{0.2}$Sr$_{0.3}$MnO$_3$,\cite{dario01a}
Nd$_{0.5}$Sr$_{0.5}$MnO$_3$,\cite{ritter00} and
La$_{0.5}$Ca$_{0.5}$MnO$_3$,\cite{huang00} the high temperature
phase is the FM one (below $T_C$), and lowering $T$ below $T_{CO}$
it turns to a CO/AFM state.

The PS is observed in compounds which are located near first order
MI transitions. In this sense, random disorder has been shown to
play a fundamental role due to its ability to overcome the small
energy difference between the competing phases near the
transition, thus inducing the inhomogeneous
state.\cite{moreo00,dario01a} Transmission electron microscopy
(TEM) studies at low $T$ in
(La$_{1-y}$Pr$_y$)$_{5/8}$Ca$_{3/8}$MnO$_3$ have shown that in the
CO/AFM matrix there exist FM clusters of irregular shape with a
size of several hundred of nanometers.\cite{uehara99} These
observations make unlikely the charge segregation in this kind of
phase separation, contrary to that observed in slightly doped AFM
manganites.\cite{allodi} The FM correlations inside a mainly
CO/AFM phase is also observed in magnetization ($M$) measurements,
which exhibit a clear FM component that allows to estimate the
overall FM phase fraction $X$.\cite{dario01a}

The first order MI transition can be tuned, for example, by
changing the average ionic radius of the A-site $\langle
r_A\rangle$. This is the case of the
Pr$_{0.65}$Ca$_{0.35-x}$Sr$_x$MnO$_3$ manganite,\cite{dario01b}
which for $x=0$ is CO/AFM and for $x=0.35$ is FM metallic, while
in the $x=0.10$ composition the PS between these phases is
observed. In the (La$_{0.25}$Pr$_{0.75}$)$_{0.7}$Ca$_{0.3}$MnO$_3$
compound, Babushkina {\it et al.}\cite{babushkina00} have explored
the MI transition by partial oxygen isotope substitution.

In the present paper, we tune the MI transition by changing the
hole concentration ($n$) in
Pr$_{0.5-\delta}$Ca$_{0.2+\delta}$Sr$_{0.3}$MnO$_3$, where
$n=0.5+\delta$, around half doping ($0.46\leq n\leq 0.54$). Since
the ionic radius of the Pr$^{3+}$ and Ca$^{2+}$ ions are similar,
the advantage of this system is that $\langle r_A\rangle$ is kept
constant, so the changes in the physical properties are
intrinsically related to the variation of $n$. We show that almost
the complete phase diagram of these samples is phase separated
below $T_C$. In the whole temperature range the amount of FM phase
continuously decreases with increasing $n$. In addition, the low
temperature MI transition of the $n\leq0.5$ samples is of
percolative nature, i.e. for $T_{MI}<T<T_{CO}$ the resistivity
decreases with increasing $T$ due to the enlargement of $X$. In
this $T$ range the conductivity is successfully described in terms
of the usual percolation theory, though possible geometrical and
low-dimensional effects need to be considered. The CMR behavior
for percolated samples is very different to that of the
nonpercolated ones.

\section{EXPERIMENT}

Powdered samples of
Pr$_{0.5-\delta}$Ca$_{0.2+\delta}$Sr$_{0.3}$MnO$_3$ were prepared
by the nitrate decomposition route. The raw materials (CO$_{3}$Sr,
CO$_{3}$Ca and metallic Mn and Pr) were mixed in the
stoichiometric amounts and disolved in nitric acid for the
nitrates formation. Then the nitrates were heated at
800$^{\circ}$C in air for 16h for their decomposition. The
resulting powders were pressed into pellets and treated in air at
1450${^\circ}$C for 12h. Finally, the samples were cooled down to
800${^\circ}$C at 3${^\circ}$C/min and then quenched to room
temperature.

The energy dispersive spectroscopy (EDS) experiments, carried out
in a Philips 515 scanning electron microscope (SEM), indicate that
the samples have no deviations from the nominal stoichiometry
within the experimental resolution. The SEM observations also show
well connected grains, whose sizes range from $3$ to $6\mu m$.
From thermogravimetric analysis (TGA) the oxygen content was found
to be $3.000\pm0.001$.\cite{oxigPS}

Powder X-ray diffraction (XRD) data were recorded on a Phillips PW
1700 diffractometer using CuK$\alpha$ radiation and a graphite
monochromator. X-ray data at room temperature were refined by the
Rietveld method with the {\small FULLPROF} program.\cite{rietveld}

The magnetization data as a function of $T$ and $H$ were measured
between $10$ and $300K$ in a commercial SQUID magnetometer
equipped with a $5T$ magnet. Electron spin resonance (ESR)
experiments were carried out in a Bruker ESP-300 spectrometer at
34GHz (Q-band) for $100K\leq T\leq 300K$. The electrical
resistivity was measured by the usual four probe method in the
$5-300K$ temperature range with $0\leq H\leq 12T$.

\section{RESULTS AND DISCUSSION}

\subsection{X-ray diffraction}

The XRD results show that the samples mainly belong to the
orthorhombic $Pbnm$ space group. However, a minority amount of a
tetragonal phase ($I4/mcm$ space group) is also present in the
samples. In Fig. 1 we present the diffractogram of the sample
$n=0.54$, which has been satisfactorily adjusted proposing a model
with both symmetries (the small vertical bars below the XRD data
indicate the position of the Bragg reflections of both
symmetries). From the Rietveld refinements, we found that the
amount of tetragonal phase increases with increasing $n$, from
$0\%$ for $n=0.46$ to $\sim 13\%$ at $n=0.54$. In the inset of
Fig. 1 we show a blow-up of the diffractograms of the $n=0.46$,
$0.50$ and $0.54$ samples, for $2\theta$ between $46.25^{\circ}$
and $48.25^{\circ}$. In this plot, a small peak can be clearly
observed at $2\theta \approx 46.5^{\circ}$ in the $0.54$ sample,
which gradually disappears with decreasing $n$. As labeled in the
figure, this peak corresponds to the (004) reflection of the
$I4/mcm$ symmetry. On the contrary, the intensity of the larger
peak at $2\theta \approx 47.2^{\circ}$ corresponding to the (220)
reflection of the orthorhombic symmetry gradually increases with
decreasing $n$. It can be also observed that the distortion of the
majority orthorhombic phase is increased as $n$ increases, as
indicated by the larger splitting between the (220) and (004)
reflections of the $Pbnm$ space group.

Our XRD results are in agreement with the previous ones. The
Pr$_{0.7}$Sr$_{0.3}$MnO$_3$ ($\delta =-0.2$ and $n=0.30$ in our
notation) has been found to belong to the orthorhombic $Pbnm$
space group\cite{knizek92} while in the
Pr$_{0.4}$Ca$_{0.3}$Sr$_{0.3}$MnO$_3$ compound ($\delta =0.1$ and
$n=0.60$) the orthorhombic phase was found to coexist with a small
amount of the tetragonal one.\cite{barnabe01}

\subsection{Magnetization and ESR measurements}

The magnetization as a function of $T$ was measured at $H=100Oe$
in zero-field cooling (ZFC) and field cooling (FC) processes, and
is shown in Fig. 2 for all the samples studied. Although all the
curves present similar transitions (PM to FM and then to AFM on
lowering $T$), several differences are found. First at all, the
$M$ values clearly decrease a factor $\sim 15$ as $n$ increases
from $0.46$ to $0.54$. Second, qualitative differences can be
observed between the samples in the left panels ($n\leq 0.50$) and
those samples in the right ones ($n>0.50$). For $n\leq 0.50$ the
temperature where the ZFC and FC data collapse on a single curve
remarkably coincides with the maximum of $|dM/dT|$, thus we
identify that temperature as $T_C$. However, for $n>0.50$ the
$T_C$ is much higher than the $T$ where the maximum derivative
occurs. In the following of the present section we will show that
this behavior is related to the fact that the ``FM" phase for the
$n>0.50$ samples is not a true long-range-order ferromagnet, but a
coexistence with a majority PM phase is the true state. For these
samples, the $T$ of maximum derivative is an artifact generated by
the growth of the FM phase fraction with decreasing $T$ below
$T_C$. Similarly, the low-$T$ AFM state (at $T<T_{CO}$) is not a
pure phase for all the samples, and it can be anticipated that the
amount of FM component remains high in the $n=0.46$ sample, while
it decreases as $n$ increases. An additional qualitative
difference is observed for $n$ below and above $0.5$: for $n>0.5$
a second increase of $M$ is found for $T$ below $T_h\sim 100K$,
which we assume to be induced by the extra holes added into the
insulating band of the CO $n=0.50$ compound ($\delta =0.02$ and
$0.04$ for $n=0.52$ and $0.54$, respectively).

The ESR spectra at 34GHz in Fig. 3 help us to distinguish the
different magnetic contributions present in the samples. In this
figure we present the results at four representative temperatures.
In the spectra of the $n=0.54$ composition there is superimposed
the signal of a {\it dpph} marker, indicating the position of the
spectroscopic splitting factor at $g=2.00$.

Above $T_C$ [see spectra in Fig. 3(d) at $270K$], all samples
exhibit a resonance line corresponding to the PM mode, centered at
$g\approx 1.97$ (independent of $T$) with a linewidth of $\sim
600Oe$. The temperature $207K$ corresponds approximately to the
position of maximum magnetization for all the samples. At this
$T$, what can be observed is that for $n=0.52$ and $0.54$ a second
broad signal appears mounted over the still present PM line. On
the contrary, in the $n\leq 0.50$ samples the PM resonance is
almost not observed and the broad line appears alone. This second
resonance is associated with a FM mode, and exhibits very
different characteristics with respect to the PM one, namely, the
center field is shifted to lower fields due to the appearance of
an internal field, the linewidth is substantially larger, and the
line presents an important asymmetry between the maximum and the
minimum with respect to the base line. In all cases, the
temperature where the FM behavior begins to manifest coincides
with $T_C$ (defined as the collapse of the ZFC and FC curves of
Fig. 2). The different response for $n>0.50$ indicates that below
$T_C$ these samples present a coexistence of the FM and PM phases.
The onset of short-range FM order indicated by the splitting of
the ZFC and FC data was also observed in the
Pr$_{0.65}$Ca$_{0.35-x}$Sr$_x$MnO$_3$ compound.\cite{dario01b}

On lowering $T$ below the CO temperature $T_{CO}\approx 175K$ (the
temperature transition to the AFM state) the PM line is totally
lost in all the samples. At $T=150K$ the only observed response
corresponds to the FM signal, which remains present also at
$100K$. The presence of this signal indicates that below $T_{CO}$
all the samples are phase separated, exhibiting the coexistence of
the CO/AFM and FM phases.

In the spectra at $T=100K$ it must be noted that the $n>0.50$
samples present an additional line [labeled with an asterisk in
Fig. 3(a)] at low fields $H\sim 2.5kOe$. The appearance of this
resonance coincides with the increase of $M$ below $T_h$. This
resonance is likely not related to the FM phase, but it must occur
within the CO volume. However, at this moment we cannot
distinguish whether it is related to a homogeneous canting of the
spins of the CO/AFM phase inducing a weak FM (WF) component or it
corresponds to a third different (FM) phase. In spite of this, we
note that WF modes in other materials have been found to appear at
low fields.\cite{fainstein93}

In order to quantify the amount of FM phase we measured $M(H)$
curves. The temperature dependent zero-field FM phase fraction
$X(T)$ can be estimated from the $T$ dependence of the spontaneous
magnetization $M_0$ [the back-extrapolation of $M(H)$ to $H=0T$].
To that end, we have measured for all the samples several $M(H)$
curves at different temperatures below $T_C$. We show these data
in Fig. 4. To obtain these curves, the samples were previously
heated up to the PM phase, then were cooled in zero field down to
$10K$ and finally heated again to the measurement temperature.
Therefore, the $X(T)$ data that we obtain from these measurements
correspond to a ZFC and increasing $T$ experiment.

The behavior observed in the $M(H)$ data of Fig. 4 even reinforce
the features described above. It becomes clear in these curves
that, although at $T_{CO}\approx 175K$ a FM to CO/AFM transition
occurs, at $10K$ a FM component still persists. This FM component
is another signature of the PS. It is also absolutely clear that
at $10K$ none of the samples are totally FM since the saturated
magnetization $M_S(n)=(4-n)\mu _B$ is never reached.

The $n=0.46$ sample, with the highest $X$ exhibit FM-like curves.
The sample with $n=0.50$, below $T_{CO}$ presents a pronounced
hysteresis in the $M(H)$ curves. Initially these curves are also
FM-like though they exhibit a low magnetization as compared with
$M_S$, but above a threshold field of the order of $1$-$2$ Teslas
$M$ begins to increase rapidly. This behavior is due to the
increase of the FM volume with increasing $H$. The increase of $X$
with applied field is due to the gain of the Zeemann energy, which
induces the swelling of the FM clusters. The need of a threshold
field is probably related to the pinning of the interface between
the FM and CO/AFM volumes. At $10K$, due to the hardness of the CO
phase, fields much higher than $5T$ are needed to induce such a
metamagnetic transition.\cite{dario01a}

The samples with $n>0.50$ behave essentially similar at low $T$.
However, for $T>T_{CO}$ these samples show, together with the
small FM component, a predominant linear behavior at high fields.
The slope is not other than the PM susceptibility that coexists
with the short-range ferromagnetism. Indeed, when the temperature
crosses $T_C$ to the PM state this slope is preserved at similar
values while the FM component disappears.

From the data in Fig. 4 we obtain the zero-field $X(T)$ as in
previous works,\cite{dario01a} estimating from the Brillouin
function the spontaneous magnetization $M_{0FM}(T)$ of a totally
FM sample with the same $M_S$ and $T_C$. After that,

\begin{equation}  \label{Xfm}
X(t) = \frac{M_0(t)}{M_{0FM}(t)}
\end{equation}

\noindent where $t=T/T_C$ is the normalized temperature. Despite
of the mean-field approximation, this method correctly reproduces
the $T$ dependence of $X$.\cite{dario01a} The obtained results for
the five samples are presented in Fig. 5. This figure somehow
summarizes the conclusions of the magnetic measurements. In the
whole $T$ range $X$ decreases with increasing $n$. This result is
consistent with that expected from the XRD observations. Since the
structural distortion decreases with decreasing doping, it is
natural that the FM metallic phase is more favored at lower $n$. A
similar doping evolution of $X$ was observed by Huang {\it et
al.}\cite{huang00} in La$_{1-x}$Ca$_x$MnO$_3$ for $x$ around
$0.5$. At $10K$ the $X$ is $\approx 67\%$ for $n=0.46$ and
$\approx 4.6\%$ for $n=0.54$ (a factor $\sim 15$, as in the $M$ vs
$T$ curves of Fig. 2). At $n=0.5$ we obtain $X\approx 18\%$,
comparable to the $19\%$ and $22\%$ of Nd$_{0.5}$Sr$_{0.5}$MnO$_3$
and La$_{0.5}$Ca$_{0.5}$MnO$_3$,
respectively.\cite{ritter00,huang00} For $n=0.52$ and $0.54$, if
the increase of $M$ for $T<T_h$ were generated by a homogeneous
canting of the AFM phase, the obtained $X$ at $T<T_h$
($T/T_C<0.4$) would be slightly higher than the actual value since
the canting contribution would be included in the measured
spontaneous magnetization.

At intermediate temperatures ($T/T_C\sim 0.7$), the CO/AFM to FM
transition is seen as an important increase of $X$. At high $T$,
the $n\leq 0.50$ samples are almost totally FM and $X$ drops
abruptly to zero at $T_C$, while the other two samples reach a
maximum $X\sim 12\%$ at $T/T_C\sim 0.75$, after which the FM
component smoothly vanishes as $T$ approaches $T_C$. It is now
clear that for $n>0.50$, the variation of $X$ above its maximum is
responsible for the difference between $T_C$ and the position of
maximum derivative of the $M(T)$.

\subsection{Percolative conductivity}

In order to analyze the resistivity-FM fraction relation, we
measured $\rho$ vs $T$ curves for all the samples at three
different fields $H=0$, $2$, and $9T$, as shown in Fig. 6.
Consequently with the $X$ behavior, the resistivity increases with
increasing $n$. Even more notable, the samples with $n\leq 0.5$
present a MI transition at $T_{MI}$, while the $n>0.5$ samples
show well insulating electrical properties with a resistivity
several orders of magnitude higher. The abrupt change of $\rho$
from $n=0.50$ to $0.52$ (more than four orders of magnitude at low
$T$) is a clear indication that the metal-insulator transition
between these samples has a percolative nature, as expected for
disorder induced PS.\cite{dagottoreview} Also the MI transition as
a function of temperature is of percolative type. As observed in
Fig. 2, the samples with $n<0.5$ do not exhibit any magnetic
transition at $T_{MI}$ justifying a metallic behavior below it. On
the contrary, the abrupt fall of $\rho$ above $T_{MI}$ coincides
with the $T$ range where the magnetization begins to increase when
the system approaches the AFM to FM transition at $T_{CO}$ (the
range $0.4<T/T_C<0.7$ in Fig. 5). Therefore, in this $T$ range the
resistivity response is unmistakable related to the changes of the
FM phase fraction, i.e. at $T_{MI}$ a percolative MI transition
occurs. Between $T_{MI}$ and $T_{CO}$ the insulating-like response
of $\rho$ is given by the enlargement of the metallic paths with
increasing $T$. Below $T_{MI}$, where the FM phase fraction is
frozen, the resistivity of the $n\leq 0.5$ samples present
metallic characteristics (slightly decreases with decreasing $T$)
due to the intrinsic dependence of the metallic volume, and the
residual $\rho$ obviously decreases with increasing $X$. The fact
that the samples with $n$ above $0.5$ are always insulating
indicates that the percolation threshold is in between the $X$
values of the $0.50$ and $0.52$ samples.

In order to understand the percolative behavior, in the range
$T_{MI}<T<T_{CO}$ we obtained the conductivity ($\sigma =1/\rho$)
as a function of $X$ for $n\leq 0.50$. In this temperature range,
$\sigma$ changes orders of magnitude due to the variation of $X$,
thus the intrinsic $T$ dependence of the metallic paths can be
neglected, at least for $n=0.50$ and $0.48$. The resulting
$\sigma(X)$ data are shown in Fig. 7 (the $\sigma$ values were
obtained from the $H=0T$ warming curve, as in the $X$ vs $T$
experiment of Fig. 5). In the usual percolation theory, the
$\sigma$ vs $X$ relation for $X$ near above the percolation
threshold (the so-called critical region) is given
by\cite{kirkpatrick73}

\begin{equation}
\sigma \propto \left( X - X_C \right)^t \label{sigmapercolada}
\end{equation}

\noindent where $X_C$ is the critical percolation threshold and
$t$ is the critical exponent. The way to obtain both these
parameters is to construct a $\log \sigma$-$\log (X-X_C)$ plot,
looking for the $X_C$ which gives the best linear relation. Once
$X_C$ is obtained, the slope of this straight line is $t$. The
proportionality factor in Eq. (\ref{sigmapercolada}) is a sample
dependent constant. In the inset of Fig. 7 we present the
$\log$-$\log$ plot, where a quite fair linearity is observed and
from which we obtained $X_C=15.5(2)\%$ and $t=0.96(5)$ (the
$\sigma$ of the $n=0.50$ sample was multiplied by $2.45$ in order
to scale with the $0.48$ data). The value of $X_C$ is in very good
agreement with three-dimensional (3D) percolation, where
thresholds between $14$ and $19\%$ are
expected.\cite{kirkpatrick73,webman76} This value of $X_C$ also
gives now a quantitative support to the expectation that the
$n=0.52$ and $0.54$ samples are not percolated. The horizontal
dotted line in Fig. 5 indicates the position of $X_C$, and it is
very clear that both these samples have a $X$ below the threshold
in the whole temperature range. Therefore, the insulating state in
these samples is an intrinsic behavior of the CO phase.

The almost linear relation between $\sigma$ and $X$ ($t=0.96$) is
somewhat surprising. In manganites the critical exponents are
usually found to be between $2$ and
$4$.\cite{babushkina00,hardy01,kim00,podzorov00} The implications
of the value of $t$ will be discussed below.

In order to study the percolation physics in the presence of a
magnetic field, we have repeated the above analysis with an
applied field $H=2T$.

The magnetization of the $n=0.48$ and $0.50$ samples was measured
by increasing $T$ at $H=2T$ after a FC process, since the
$\rho(T)$ curves in Fig. 6 measured with nonzero field were
obtained by this procedure. The $M(T)$ data are shown in Fig. 8,
where all the magnetic transitions are still observed. In this
case, the obtention of the FM phase fraction requires some
considerations. On one hand, since the contribution of the AFM
phase ($M_{AFM}$) at $H=2T$ is different from zero, contrary to
the case of the zero-field (spontaneous) magnetization $M_0$, the
value of $M$ is not directly proportional to $X$. Therefore, the
$M_{AFM}$ needs to be subtracted. On the other hand, when
comparing with the magnetization of a totally FM sample, it must
be done at the field $H=2T$. However, this point can be neglected
since for $T$ below $0.75T_C$, in a totally FM material the
difference between the magnetization at $H=2T$ and the $M_{0FM}$
is negligibly small. For example, in the $n=0.46$ sample which at
$T=200K$ ($0.75T_C$) has a zero-field $X\approx 91\%$, the $M_0$
is $\approx 2.25\mu _B$ while the $M$ at $H=2T$ is $\approx
2.34\mu _B$, i.e. the difference is $\approx 4\%$, even including
a minor contribution of the AFM volume and a possible
field-induced increase of $X$. Finally, with the above
considerations we can write

\begin{equation}
M \approx X M_{0FM} + (1-X) M_{AFM} \label{2Tfraction}
\end{equation}

The $M_{AFM}$ can be obtained from the parent compound
Pr$_{0.5}$Ca$_{0.5}$MnO$_3$, which is totally CO/AFM. Indeed, this
compound exhibits well linear $M(H)$ curves with no spontaneous
magnetization.\cite{krupicka99} In the $T$ range of our interest
($75$-$175K$), the susceptibility of this compound is weakly
temperature dependent, lying between $0.040$ and $0.045\mu
_B/T$,\cite{krupicka99} which indicates that $M_{AFM}\approx
0.08$-$0.09\mu _B$. After this contribution was taken into
account, the $X$ at $H=2T$ was calculated from Eq.
(\ref{2Tfraction}) and the $\sigma$ vs $X$ curve of Fig. 9 was
obtained. In this case, to obtain the scaling with the $n=0.48$
sample, the conductivity of the $n=0.50$ one was multiplied by
$1.8$. The $\log$-$\log$ plot of these data (inset of Fig. 9) also
exhibits a good linear correlation with $X_C=17.8(2)\%$ (also in
agreement with 3D percolation), and the slope gives an exponent
$t=1.21(3)$, slightly higher than that without applied field. For
comparison, in the same inset we have also included the straight
line with slope $t=0.96$ (dotted line) in order to highlight the
increase of the exponent with the magnetic field.

\subsection{Critical exponents: Relation to percolation mechanisms}

Regarding the critical exponent, some experimental works have been
performed in PS manganites. In some of them,\cite{kim00} the
authors used the exponent $t=2$ expected for a mean-field behavior
of the 3D percolation problem\cite{kirkpatrick76} and deduced
$X_C$. In other cases,\cite{babushkina00,hardy01} as in this paper
both $X_C$ and $t$ were deduced, and the obtained exponents lie in
the range $2$-$4$. However, these works were done at doping levels
between $0.3$ and $3/8$, while our samples are situated around
half doping. It is feasible that the precise percolation mechanism
would depend on the doping level.

There are several ways to change the value of the critical
exponent. A lot of experimental and theoretical work focused on
this problem indicate that for 3D percolation, $t$ lies between
$1.5$ and $2$.\cite{kirkpatrick73,webman76,kirkpatrick76,coutts76}
However, it has been shown that the conductivity exponent is
nonuniversal,\cite{lee86} and specific geometrical configurations
of the contacts between percolating clusters could shift $t$ to
much higher values, e.g. by the formation of narrow necks. This is
in agreement with the exponents found at the $0.3$-$3/8$ doping
range, but in our case $t$ is well below $1.5$.

There are other ways to produce a decrease on the critical
exponent. One of them is also related to the geometry of the
percolating clusters. It must be taken into account that the
important parameter is not the conductivity, but the percolation
probability $P(X)$ must be considered.\cite{kirkpatrick73} This
probability represents the fraction of the sample which belongs to
a percolated metallic path (thus $P(X)\leq X$). The percolation
probability also has a critical behavior as in Eq.
(\ref{sigmapercolada}) with an exponent $s$ which is typically
$\sim 0.5$ for common lattice percolation
models\cite{kirkpatrick73} and $s=1$ for other sophisticated
branching models which consider Bethe
lattices.\cite{stinchcombe73} Then, the conductivity can be
obtained as\cite{kirkpatrick73,stinchcombe73}

\begin{equation}
\sigma (X) = \mu (X) P(X) \label{movilidad}
\end{equation}

\noindent where $\mu$ is an effective mobility determined by the
shape of the percolated (infinite) cluster. Therefore, the
percolation probability $P(X)$ is an upper bound for the
conductivity. The optimal case would occur if the connected region
were arranged in regular arrays parallel to the applied electric
field. This would result in the maximum $\mu$, and the ratio of
the conductivity of the whole material to the bulk conductivity of
the metallic part would be equal to $P(X)$,\cite{kirkpatrick73}
then $t=s$. Indeed, the asymmetry observed in the FM signal in our
ESR experiments could be showing some degree of shape anisotropy
of the FM clusters. On the other hand, the intricate and
constricted nature of the percolated paths reduce the mobility.
For predominantly chain-like infinite clusters\cite{stinchcombe73}
$\mu \sim (X-X_C)$, thus $t\simeq s+1$. As a conclusion, the role
of the precise topology of the percolated clusters gives rise to a
variety of nonuniversal exponents.

On the other hand, exponent values in the range $1$-$1.4$ are
naturally found in two-dimensional percolation
systems.\cite{kirkpatrick73,coutts76,dubson85} This seems to
contradict the fact that $X_C$ remarkably coincides with the 3D
threshold, however it must be taken into account that near half
doping the mixed-valence manganites exhibit tendencies toward 2D
metallic states in the AFM phase. For example
La$_{1-x}$Sr$_x$MnO$_3$ for $x\gtrsim 1/2$,
Pr$_{0.5}$Sr$_{0.5}$MnO$_3$, and Nd$_{0.45}$Sr$_{0.55}$MnO$_3$ all
present the A-type AFM structure at low
$T$.\cite{jirak00,moritomo98,dario00,kuwahara99} In this magnetic
phase the $ab$ planes are ferromagnetically ordered, while
contiguous layers have opposite spin direction. In the A-type
phase, the electronic properties have been associated to a 2D
metallic state due to a $d_{x^2-y^2}$ orbital ordering within the
FM planes.\cite{dario00,kuwahara99,maezono98} It is possible that
in our Pr$_{0.5-\delta}$Ca$_{0.2+\delta}$Sr$_{0.3}$MnO$_3$ samples
part of the AFM volume would be of the A-type, exactly in the same
way as in Nd$_{0.5}$Sr$_{0.5}$MnO$_3$, where low $T$ neutron
powder diffraction (NPD) experiments show that $20\%$ of the
sample belongs to this phase.\cite{ritter00} Indeed, our samples
share a common characteristic with the three previously mentioned
A-type compounds. Those manganese perovskites present at room
temperature a tetragonal crystal
structure,\cite{jirak00,moritomo98,dario00,kuwahara99} similar to
the small amount of the $I4/mcm$ space group present in our
samples. In fact, Woodward {\it et al.}\cite{woodward98} suggested
that near half doping, this tetragonal space group favors the
occupancy of the $d_{x^2-y^2}$ orbitals with the A-type AFM phase
as ground state. The tetragonal volume in our samples should then
produce the appearance of A-type AFM droplets at low $T$. Of
course our magnetization measurements cannot distinguish between
the CE-type (of the CO phase) and the A-type AFM states.

It could be possible that near the percolation threshold of the FM
phase, some intercluster regions of the A-type connect the
clusters as a sort of {\it valve} effect producing an effective 2D
behavior, although the value of $X_C$ determined by the
dimensionality of the clusters themselves is unaltered. When a
magnetic field is applied, the spins of the A-type AFM structure
begin to orient in the field direction, with a consequent
improvement of the electronic coupling between the otherwise
decoupled metallic planes. This produces a field-induced
dimensional crossover\cite{kuwahara99} from 2D to 3D that could be
responsible for the increase of $t$ from $0.96$ to $1.21$.

The above discussion makes totally clear the need of a close
understanding of the percolation mechanism in phase-separated
manganites. In this sense the geometry and effective
dimensionality as well as doping evolution of the FM clusters
become relevant.

\subsection{Magnetoresistance}

A notable feature in the $\rho (T)$ curves of Fig. 6 is that for
$n\leq 0.50$ an appreciable CMR is obtained with a moderated field
of $2T$, while for $n>0.50$ at this field the resistivity remains
almost unchanged. In the latter case, since the FM phase is not
percolated the resistivity is dominated by the intrinsic behavior
of the CO phase. A proof of this is given by Fig. 10, where we
present MR vs $H$ curves at $T=50K$. For the percolated samples
($n\leq 0.50$), at the first stages of the $\rho$ vs $H$ loops
($H<0.5T$) there is clearly visible a rapid decrease of $\rho$
with increasing $H$, and after the field cycling (that produces a
considerable hysteresis) the resistivity notably returns to its
original value at $H=0T$. The initial drop of $\rho$ at low fields
is related to the LFMR of polycrystalline FM
manganites,\cite{hwang96} induced by the spin-polarized tunneling
of electrons traveling through percolated paths that cross the
grain boundaries. On the other hand, since the $n>0.50$ samples
are not percolated, there are no metallic paths connecting the
electrodes across the whole sample. As a result, the rapid
decrease of $\rho$ at low fields is not observed. On the contrary,
a smooth decrease occurs at low fields, typical of CO manganites.
In addition, when the magnetic field is removed in the MR curves
of Fig. 10(b), the resistivity does not go back to the initial
value, but a much higher conductivity is retained. However, the
insulating-like $\rho (T)$ curves in Figs. 6(d) and (e) indicate
that, even in a FC process fields well above $9T$ are necessary to
induce the percolation of the FM metallic phase in the $n>0.50$
samples.

In order to show the field evolution of the $X$ dependence of the
MR, in Fig. 11 we show the zero-field $X(n)$ curve at $T=50K$ in
the top panel and in the bottom one the ratio $\rho(H) /\rho (0)$
vs $n$. It is clear that at low fields, the percolated samples
exhibit the more important MR values. As the field increases above
$1T$, however, the increase of the FM volume provides the dominant
contribution producing better MR values in the $n=0.50$ sample. As
the field reaches much higher values, the relative change of
resistivity shifts the higher MR values to the nonpercolated
compounds (see the $X_C$ line in the top panel).

The results shown in Fig. 11(b) indicate that the ideal materials
to produce the highest MR values should be that located close to
the percolation threshold. It would be even better if the
zero-field FM phase fraction is located slightly below $X_C$, in
such a way that a small applied field produces the increase of FM
volume necessary to percolate the metallic phase thus raising the
conductivity several orders of
magnitude.\cite{mayr01,babushkina99}

\subsection{Phase diagram}

As a summary of our results, in Fig. 12 we present a complete
zero-field $T$-$n$ phase diagram. At high temperature, the $M$ and
ESR measurements show that the samples are in the PM phase. On
lowering $T$ below $T_C$, FM correlations appear in all the
samples. Although the $T_C$ is symmetric around $n=0.50$, in the
samples with doping above this value the FM phase is not
long-range-ordered, but coexists with a majority PM phase. In
agreement with this, the $\rho (T)$ curves do not show any hint at
$T_C$, while for $n<0.5$ the resistivity presents a local maximum
at that temperature.

At the CO temperature $T_{CO}$ the samples turn into the CO/AFM
state. However, a fraction of FM phase persists in all the samples
down to the lowest temperatures. The region immediately below
$T_{CO}$ is characterized by insulating resistivities, however the
electrical properties are again different depending on the doping
level. For $n>0.50$, due to the low values of $X$ the insulating
state is given by an intrinsic response of the CO volume. On the
contrary, for $n\leq 0.50$ the decrease of the FM volume fraction
with decreasing $T$ promotes the insulating behavior, though there
are percolated metallic paths. In these samples, when the FM phase
fraction is frozen a MI transition occurs at $T_{MI}$, below which
the $T$ evolution of $\rho$ is intrinsic of the metallic phase.
Between $T_{MI}$ and $T_{CO}$ is where the usual percolation
theory accounts for the conductivity variation.

The temperature $T_h$ for $n>0.50$ indicates the additional
magnetization increase at low $T$ with the simultaneous appearance
of a low-field resonance line in the ESR spectra. Although this is
labeled in the phase diagram as the $F'$ phase, at this moment we
cannot distinguish whether this contribution comes from a third
phase or from a homogeneous canting of the AFM state. NPD
experiments are being performed to elucidate the origin of this
magnetic component. These experiments will also help to probe
whether or not there exists a small A-type AFM volume in the
samples studied here.

By extrapolating the FM fraction toward lower holes concentration,
we estimate that for $n\sim 0.43$ the material should be $100\%$
FM. At this doping level the $T_{MI}$ and $T_{CO}$ should collapse
at the same value, and the compound should present metallic
properties in the whole $T$ range below $T_C$. The
Pr$_{0.6}$Ca$_{0.1}$Sr$_{0.3}$MnO$_3$ ($n=0.40$) and the limiting
compound Pr$_{0.7}$Sr$_{0.3}$MnO$_3$ ($n=0.30$) are indeed in this
situation.\cite{lin01} For $n>0.50$ the FM phase is strongly
suppressed and the material is essentially CO, in agreement with
the much higher structural distortion. In the intermediate region,
where the first order phase transition occurs between the metallic
and insulating phases the PS is observed, where $X$ smoothly
decreases with increasing $n$ producing a percolative MI
transition.

\section{SUMMARY}

We studied the magnetic and electric properties of polycrystalline
Pr$_{0.5-\delta}$Ca$_{0.2+\delta}$Sr$_{0.3}$MnO$_3$ around half
doping. The magnetization and ESR measurements allowed us to
characterize the magnetic states of the compound in the whole $T$
range, as well as to obtain the FM phase fraction $X$ of the PS
states as a function of temperature. Consequently with the FM
phase fraction behavior, the electrical properties are
distinguishable different for $n$ below and above $0.50$, related
to the percolative metal-insulator transition as a function of
doping. In the percolated samples ($n\leq 0.50$), the increase of
resistivity with decreasing $T$ below $T_{CO}$ is related to the
reduction of the FM volume. In this $T$ range, the conductivity
follows the percolative behavior $\sigma \sim (X-X_C)^t$. The
values obtained for $X_C$ are in agreement with that expected for
conventional 3D percolation. However, the low values obtained for
$t$ depart from that observed in other compounds, indicating that
there should be different percolation mechanisms. On one hand the
geometric configuration of the FM clusters is an important issue
that needs to be addressed, and on the other hand possible 2D
effects could also produce this behavior. In this sense, the
increase of $t$ with applied field could indicate a field-induced
dimensional crossover.

With respect to the MR response, we showed that FM fractions near
$X_C$ are the optimal for producing large MR ratios with moderated
fields. In fact, the percolated samples show very different
magnetoresistance properties, specially at low fields.

Finally, the presented $T$-$n$ phase diagram shows that in a small
region of doping level around $1/2$ several electronic phases need
to be considered. The asymmetric electronic properties observed
around half doping can be regarded as related to the doping
dependence of the FM phase fraction, i.e. at half doping the
percolation of the FM metallic phase occurs.

\begin{center}
{\small {\bf ACKNOWLEDGMENTS} }
\end{center}

We thank A. Caneiro, A.A. Aligia, and J. Milano for useful
discussions and for the critical reading of this manuscript. This
work was supported by CNEA (Comisi\'{o}n Nacional de Energ\'{\i}a At\'{o}mica),
CONICET (Consejo Nacional de Investigaciones Cient\'{\i}ficas y
T\'{e}cnicas), Fundaci\'{o}n Antorchas and ANPCyT (Agencia Nacional de
Promoci\'{o}n Cient\'{\i}fica y Tecnol\'{o}gica, Argentina), PICT 99-03-05266.
RDS and LM acknowledge the CONICET for financial support.

\noindent $^*$ Author to whom correspondence should be addressed.
E-mail address: niebied@cab.cnea.gov.ar

\bigskip

FIG. 1. Room temperature X-ray diffractogram of the sample with
doping $n=0.54$. ($\circ$) Experimental data and (|) Rietveld
refinement using a model with two symmetries: the orthorhombic
$Pbnm$ and the tetragonal $I4/mcm$. ($|$) Bragg reflections of
both these symmetries. The lower solid line centered at $-0.25$ is
the difference between the experimental data and the refinement.
Inset: Blow-up of a selected region of the diffractograms of the
indicated samples. This shows the increase of the amount of
tetragonal phase with increasing $n$.

\bigskip

FIG. 2. Magnetization vs temperature at $H=100Oe$ in ZFC (lower
curves) and FC (upper curves) processes. The doping level is
indicated in each case. The $T_C$ is the temperature where the ZFC
and FC data collapse on a single curve.

\bigskip

FIG. 3. ESR spectra obtained at $34GHz$ for all the samples (as
labeled) at four representative temperatures indicated in the
panels. The narrow line mounted over the spectra for $n=0.54$
corresponds to the {\it dpph} marker, which indicates the value
$g=2.00$. The asterisks in panel (a) show the additional line that
appears at the temperature $T_h\sim 100K$.

\bigskip

FIG. 4. $M(H)$ curves at several $T$ for the four samples
indicated. The sample with $n=0.48$ (not shown) presents an
intermediate behavior between $n=0.46$ and $0.50$. All these
curves were obtained after heating the samples up to the PM phase,
followed by a ZFC to $10K$ and subsequent heating to the $T$ of
the measurement.

\bigskip

FIG. 5. Zero-field FM phase fraction $X$ as a function of the
reduced temperature. The samples with $n>0.50$ show a small FM
volume which does not percolate, while for $n\leq 0.50$ the high
$X$ values lie above the percolation threshold $X_C$ (dotted
line).

\bigskip

FIG. 6. $T$ dependence of the resistivity for all the samples at
several applied fields. Those curves measured with $H\neq 0T$ were
obtained in FC process. The arrows indicate the direction of the
temperature variation and $T_{MI}$ the metal-insulator transition
temperature for the $n\leq 0.50$ samples.

\bigskip

FIG. 7. Conductivity vs FM phase fraction ($X$). The data of the
$n=0.50$ sample were multiplied by $2.45$. The line is a fit with
the percolation law $\sigma \sim (X-X_C)^t$ Inset: $\sigma$ vs
$(X-X_C)$ in $\log$-$\log$ scales, showing the percolative
behavior with the indicated values of $X_C$ and the exponent $t$.

\bigskip

FIG. 8. $M$ vs $T$ for $n=0.48$ and $0.50$ measured with an
applied field $H=2T$ and increasing temperature after a FC
process.

\bigskip

FIG. 9. $X$ dependence of $\sigma$ at $H=2T$. In this case the
$\sigma$ for $n=0.50$ was multiplied by $1.8$. The data are well
fitted with Eq. (\ref{sigmapercolada}) using $X_C=17.8\%$ and
$t=1.21$ (solid line). Inset: $\log \sigma$ vs $\log (X-X_C)$,
where a linear dependence characteristic of a percolation process
is observed. The straight dotted line corresponds to an exponent
$t=0.96$, obtained at zero-field.

\bigskip

FIG. 10. Magnetoresistance vs $H$ at $T=50K$ of all the studied
samples, as labeled. The arrows indicate the direction of the
field variation.

\bigskip

FIG. 11. a) $X$ (at $H=0T$) as a function of $n$ at $T=50K$. The
dotted line indicates the position of the percolation threshold.
b) MR vs $n$ at the same $T$ for several applied magnetic fields,
as labeled.

\bigskip

FIG. 12. Phase diagram for the polycrystalline
Pr$_{0.5-\delta}$Ca$_{0.2+\delta}$Sr$_{0.3}$MnO$_3$ (hole doping
$n=0.5+\delta$). Except for $T>T_C$ and between $T_C$ and $T_{CO}$
for $n<0.50$, the remaining diagram is completely phase separated.
Below $T_{CO}$ the CO and FM phases coexist, but the gray region
correspond to the percolated samples which follow the indicated
percolation law. Below the $T_{MI}$ the FM fraction is frozen and
the metallic paths show their intrinsic $T$ dependence. For
$n>0.50$, the $T_h$ indicates the appearance of the additional
resonance line in the ESR spectra concomitant with the
magnetization increase. Though we call this the $F'$ phase, we
cannot distinguish between an additional FM phase and a
homogeneous canting of the spins in the AFM volume.

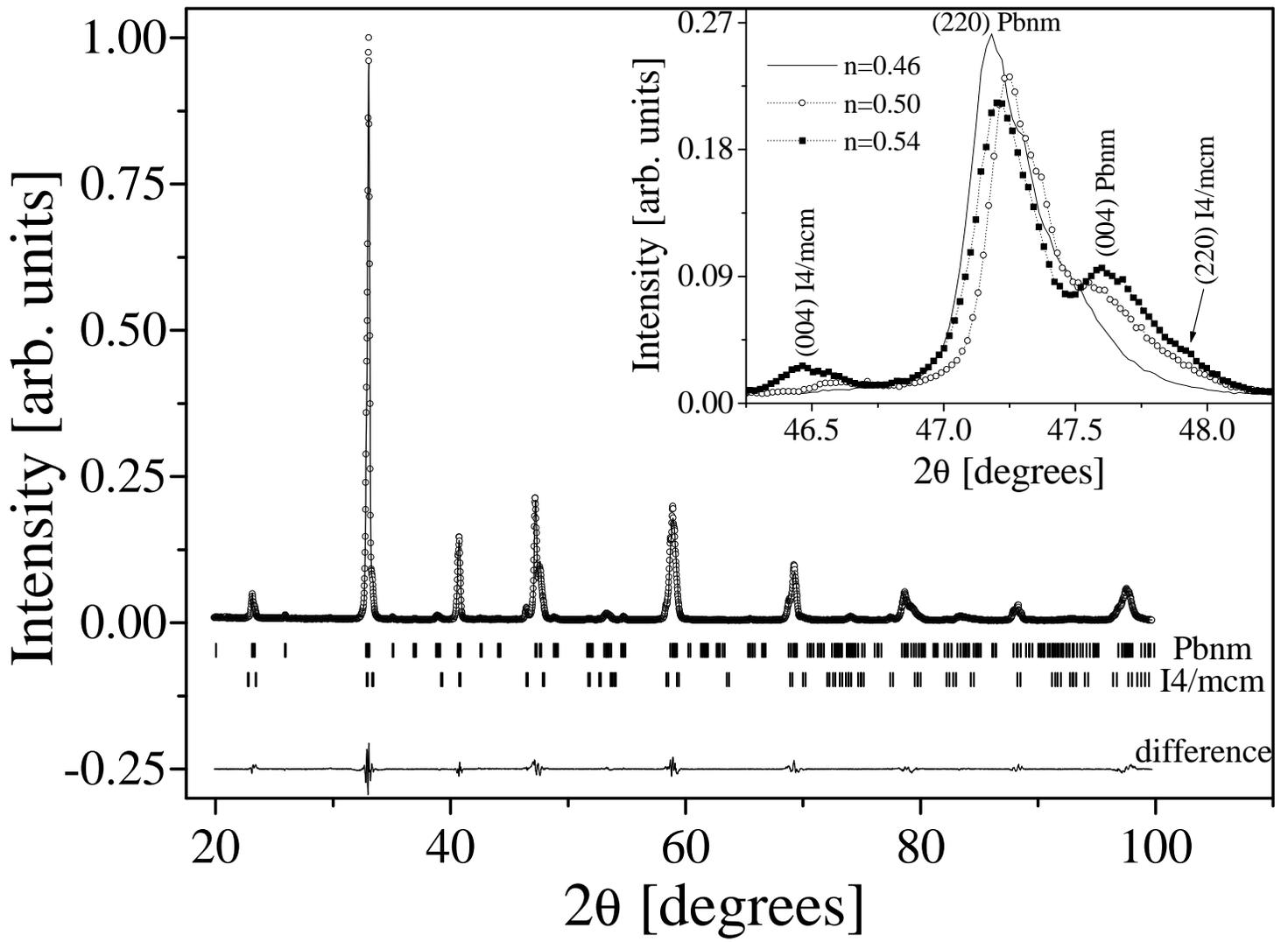
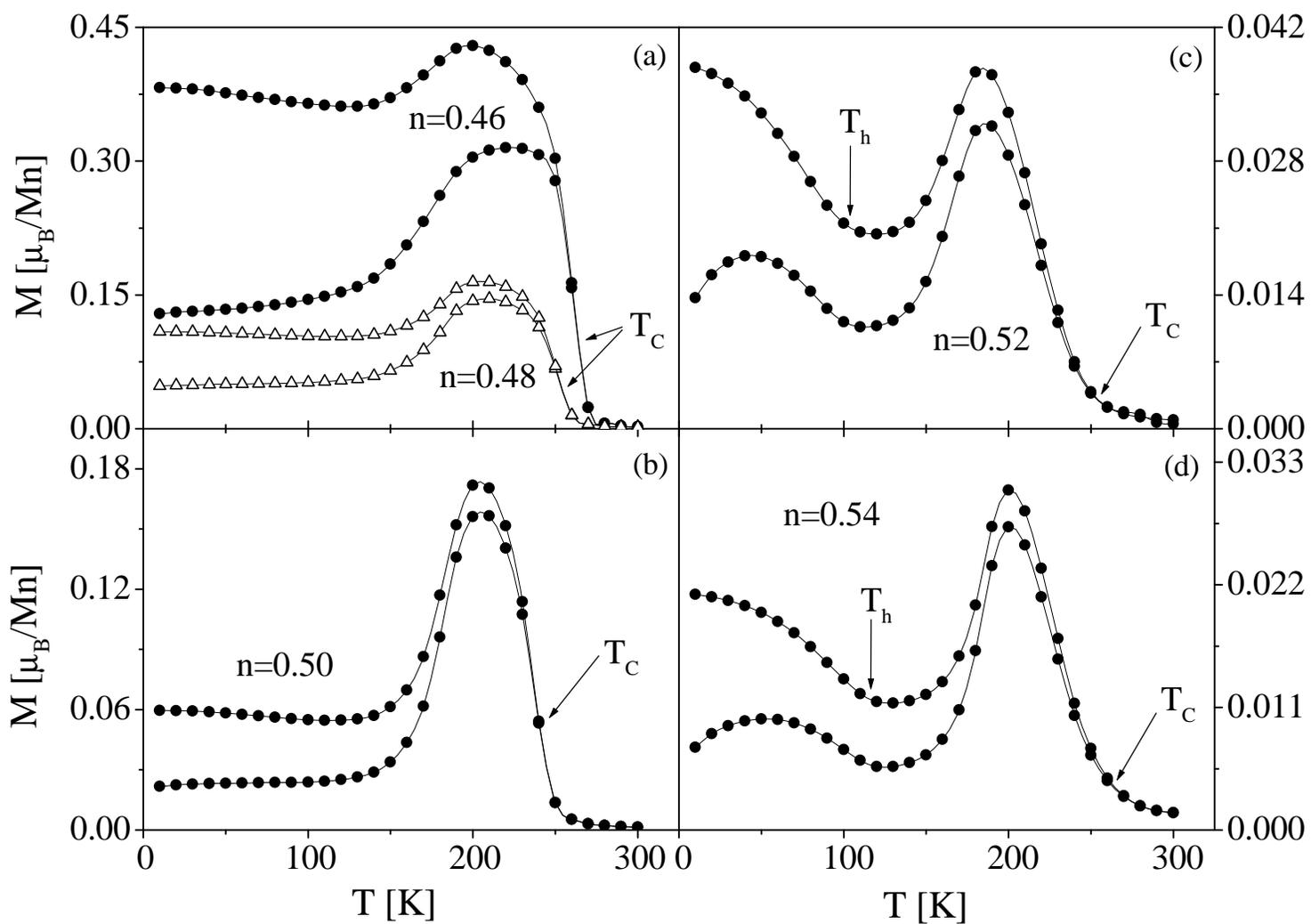
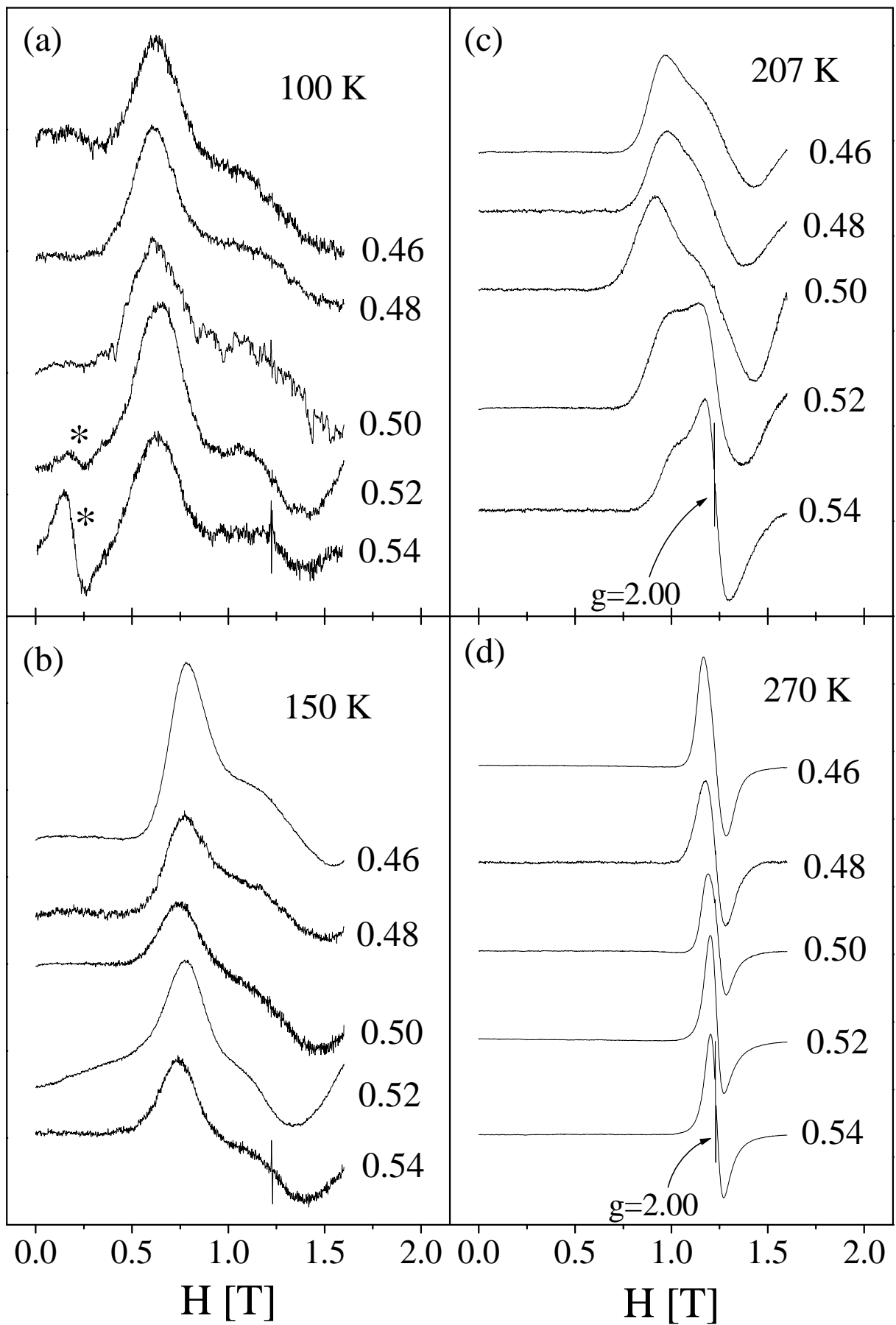
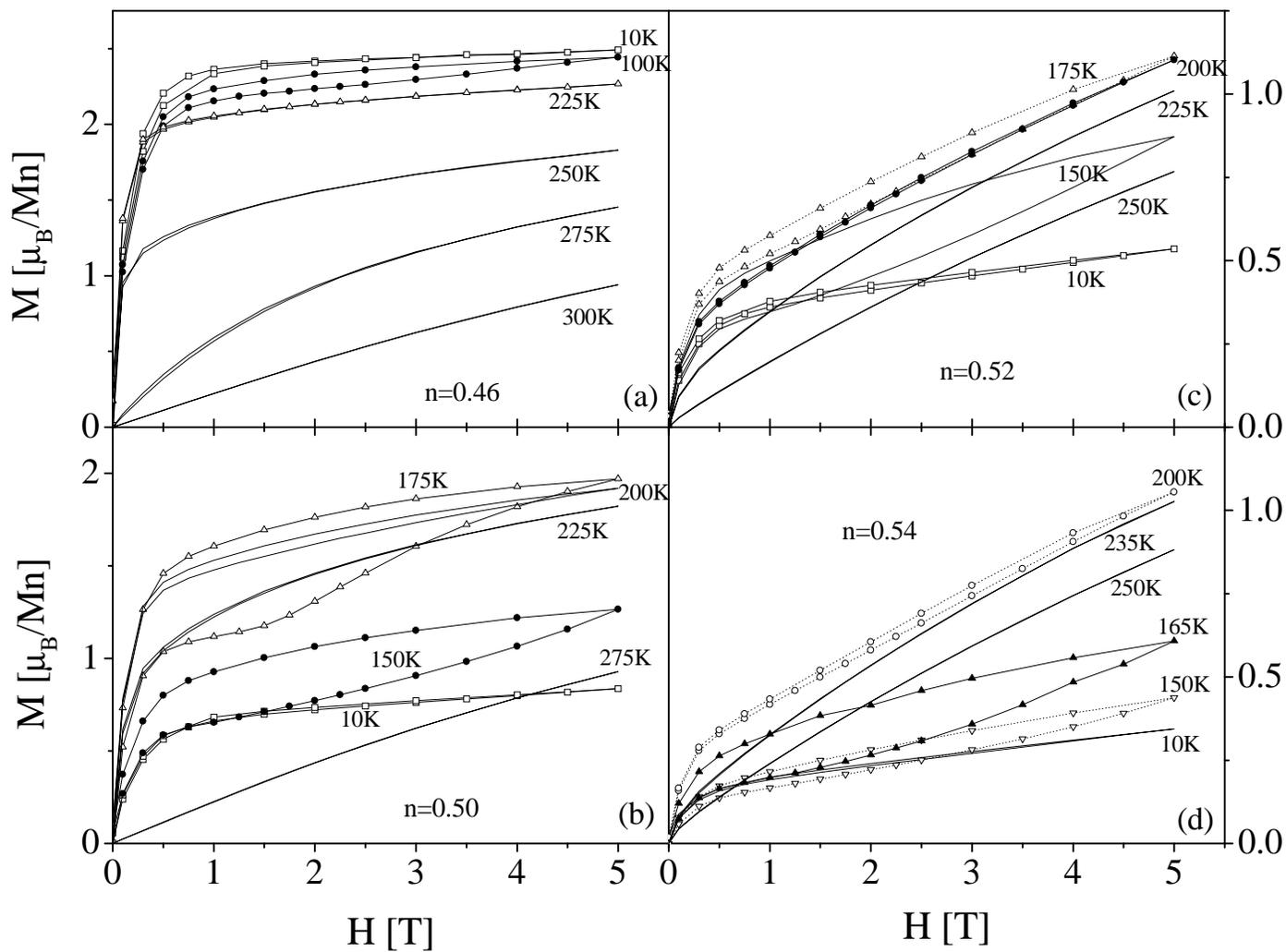
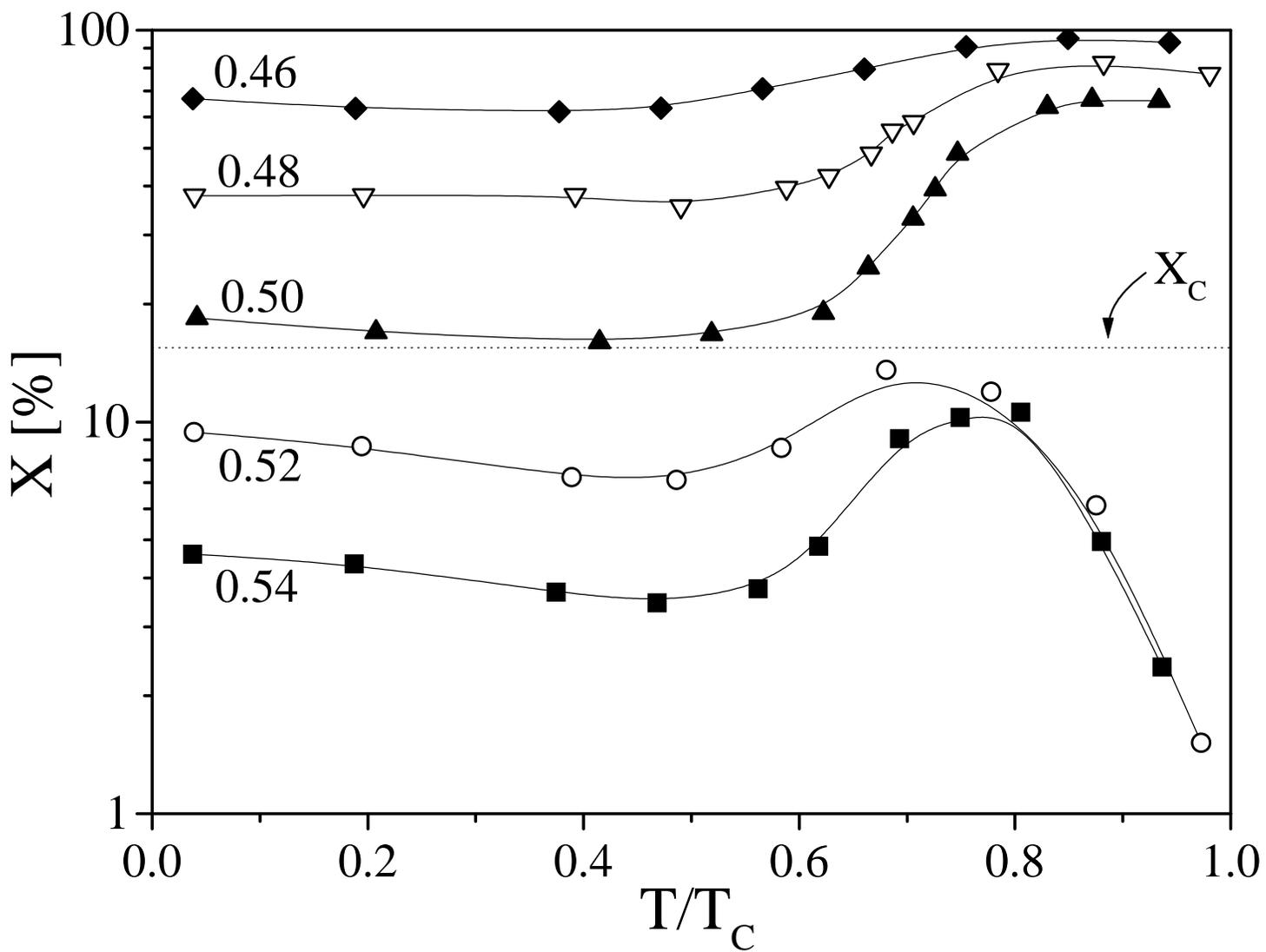
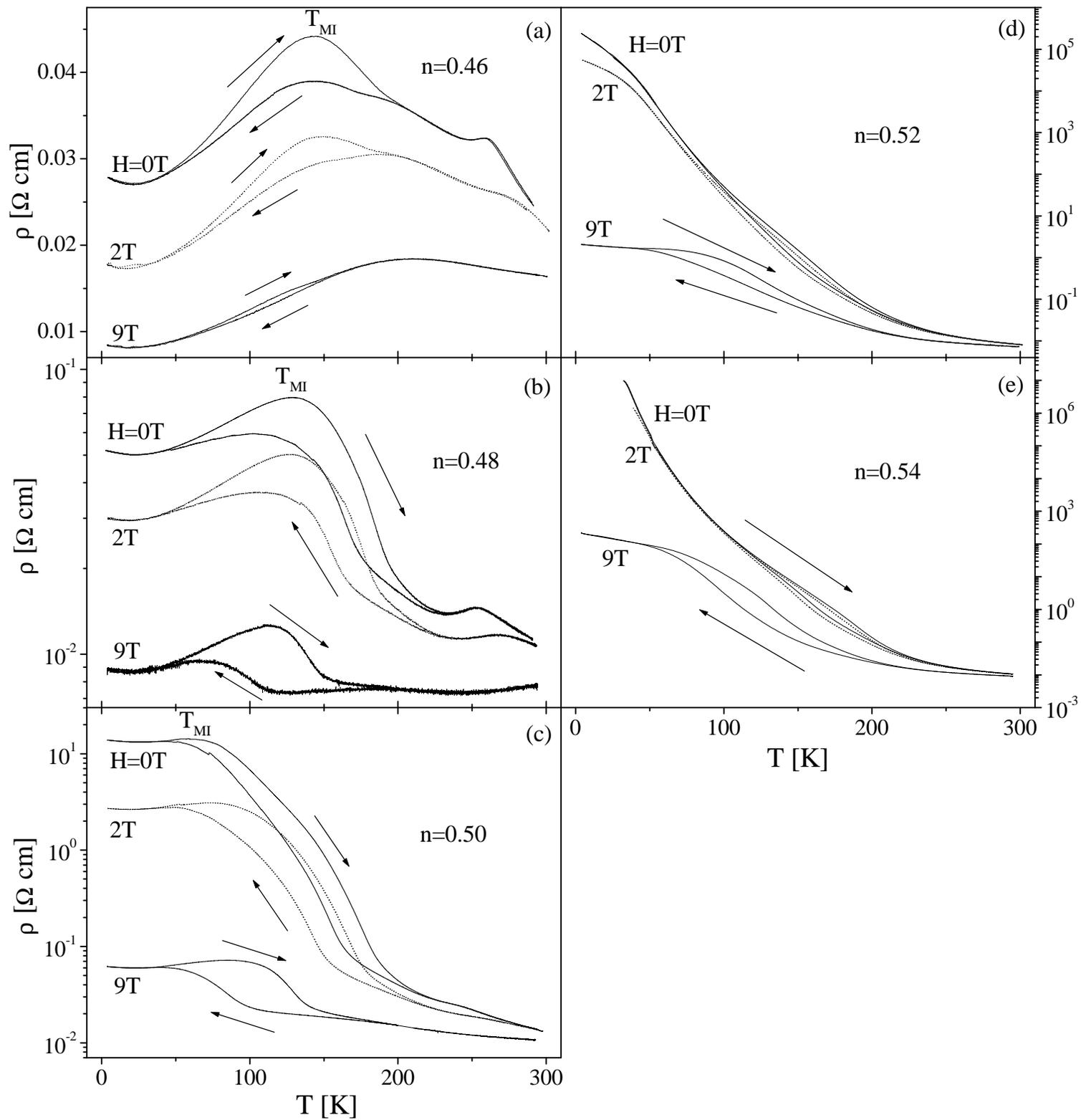
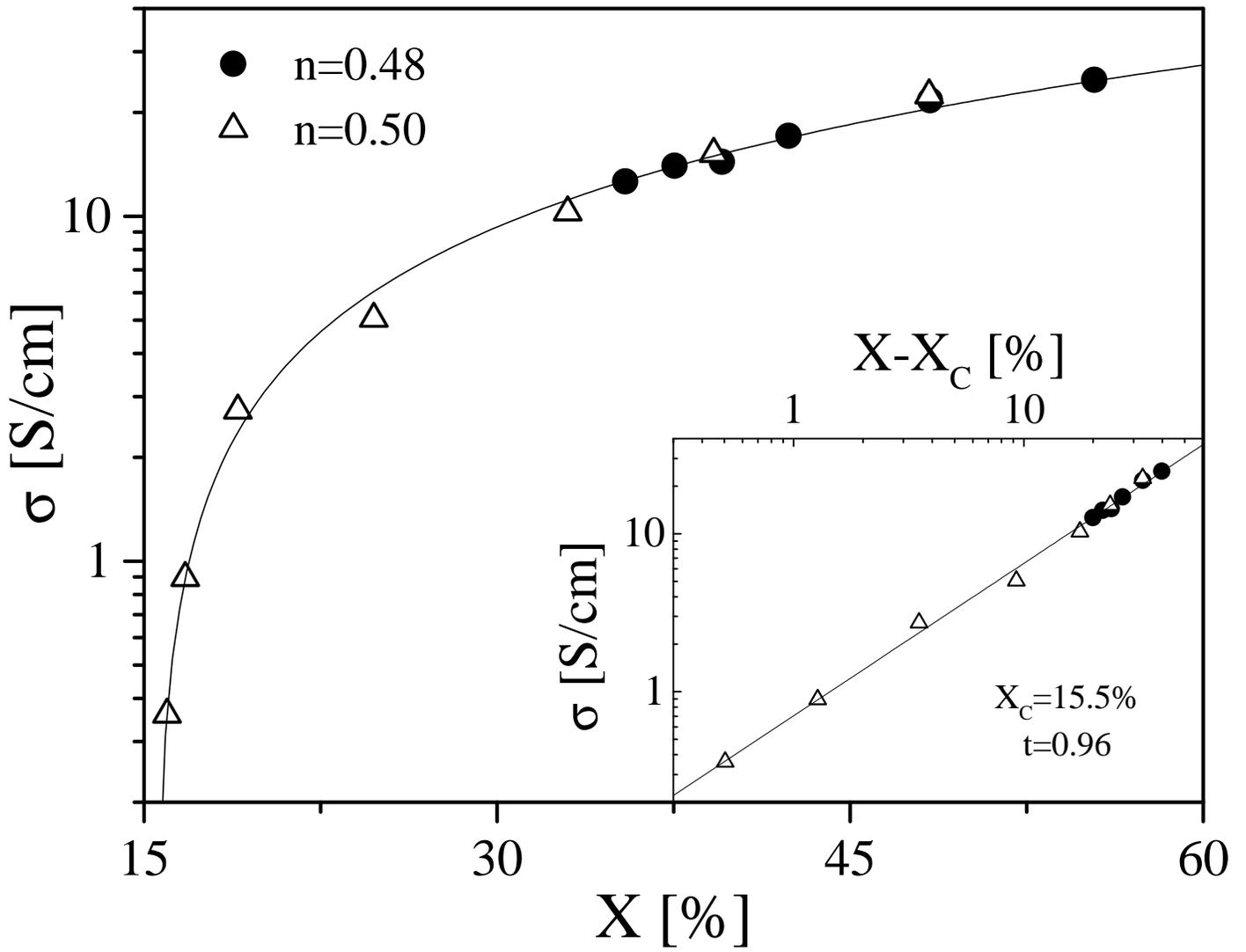
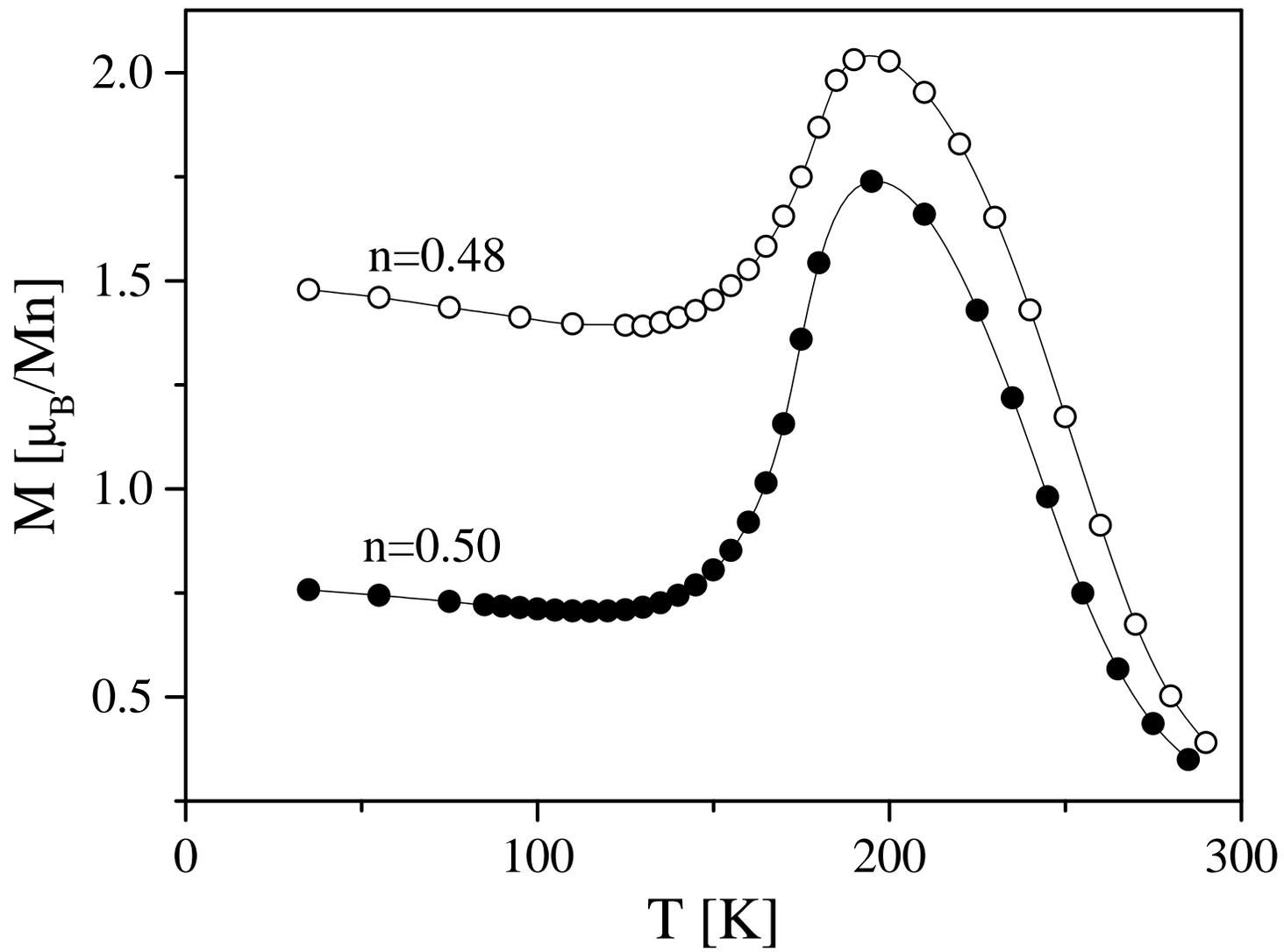
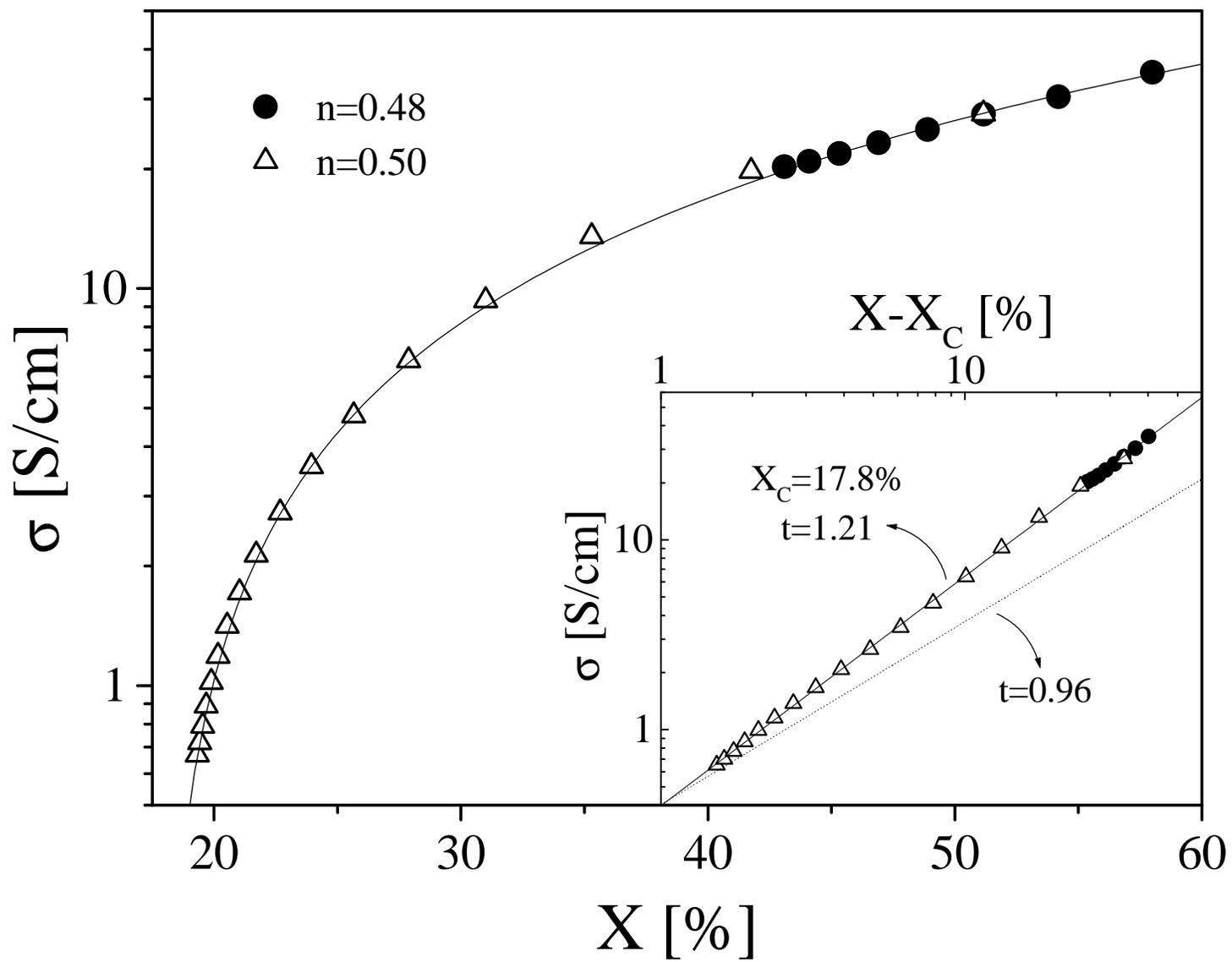
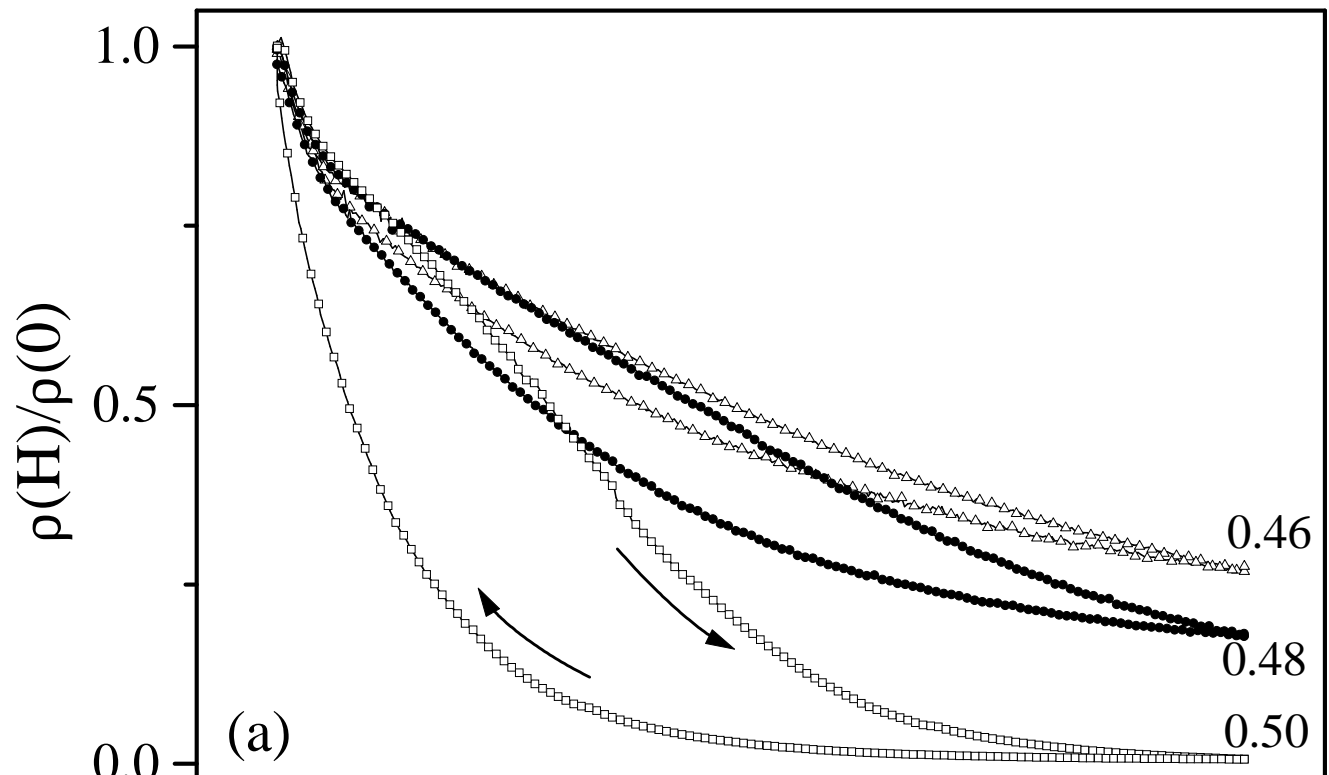
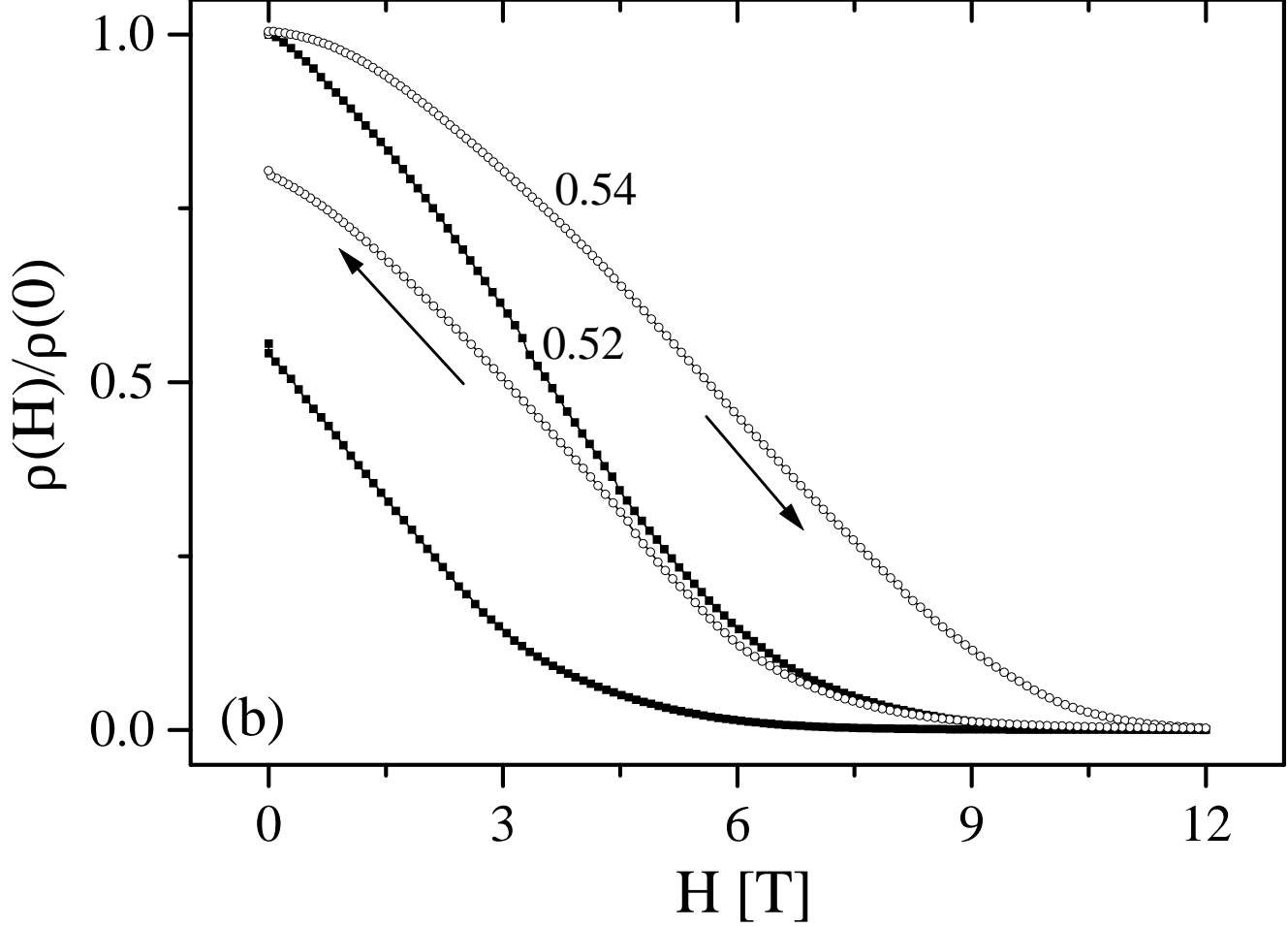
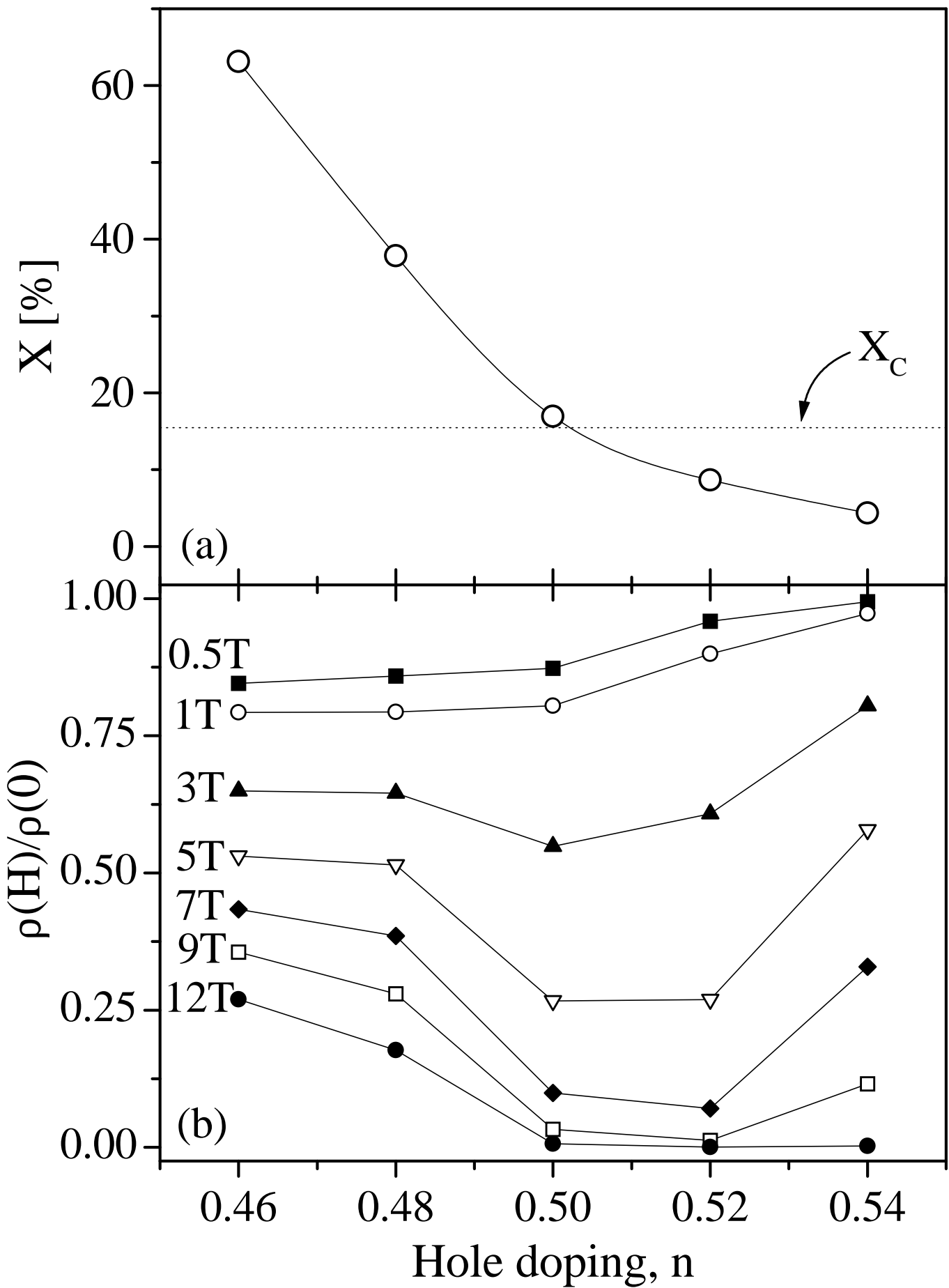
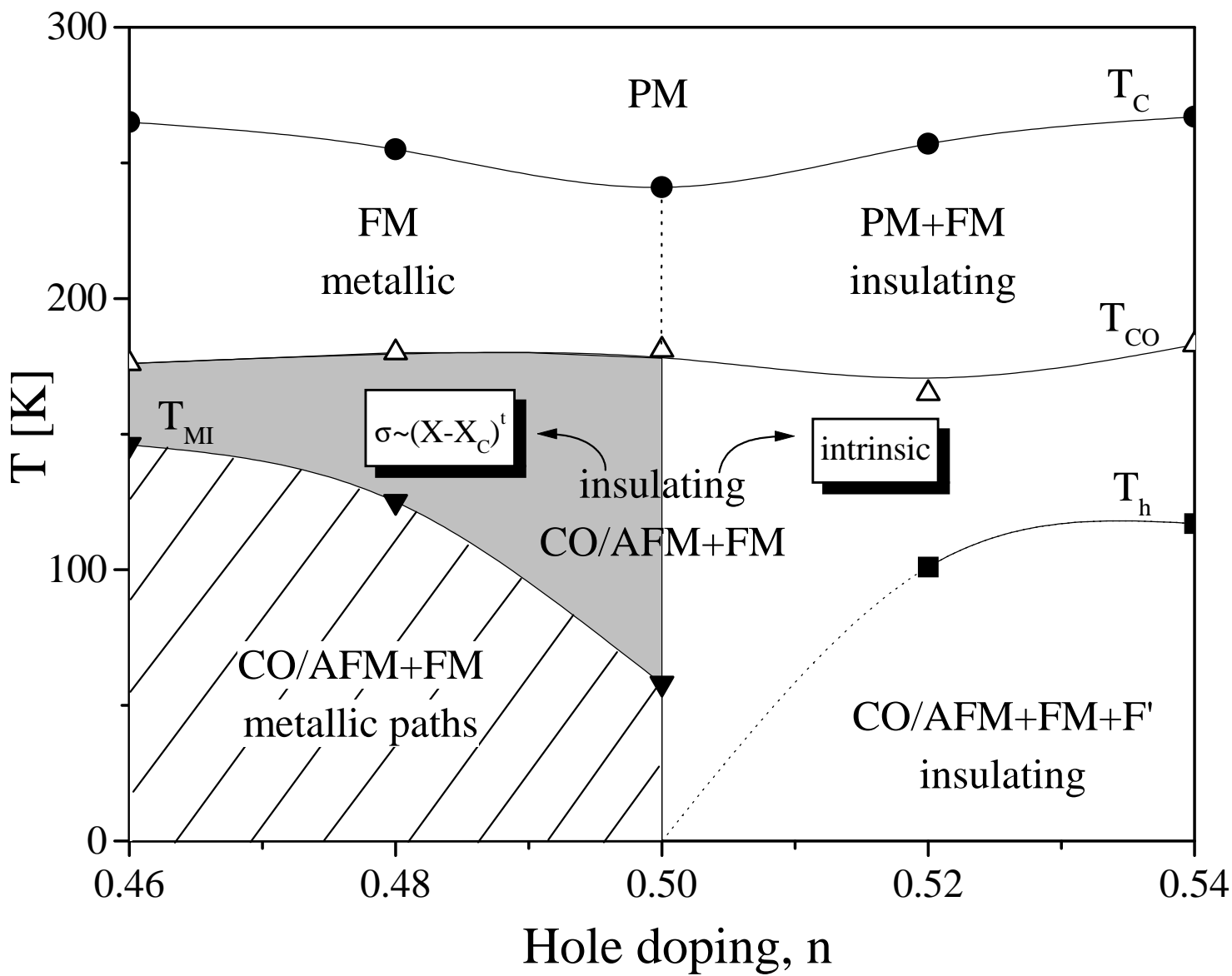
\end{document}